\documentclass[preprints,conferencereport,accept,moreauthors,pdftex,10pt,a4paper]{mdpi}

\firstpage{1}
\pubvolume{6}
\issuenum{0}
\articlenumber{00}
\pubyear{2018}
\copyrightyear{2018}
\history{Received: 2 November 2018; Accepted: 21 November 2018; Published: date}

\updates{yes}

\usepackage{amssymb}
\newcommand{\dd}{\mathrm{d}}

\Title{Magnetizing the Cosmic Web during Reionization}

\Author{Mathieu Langer $^{1,}$*$^{,\dagger}$\orcidA{} and Jean-Baptiste Durrive $^{2,}$*$^{,\dagger,\ddagger}$}
\AuthorNames{Mathieu Langer and Jean-Baptiste Durrive}

\address{%
$^{1}$ \quad Institut d'Astrophysique Spatiale, CNRS/Universit\'e Paris-Sud, Universit\'e Paris-Saclay, b\^atiment 121, Universit\'e Paris-Sud, 91405 Orsay Cedex, France\\
$^{2}$ \quad Department of Physics and Astrophysics, Nagoya University, Nagoya 464-8602, Japan}

\corres{Correspondence: mathieu.langer@ias.u-psud.fr (M.L.); jdurrive@irap.omp.eu (J.-B.D.)}
\firstnote{These authors contributed equally to this work.}
\secondnote{Current address: Institut de Recherche en Astrophysique et Plan\'etologie (IRAP), CNRS, Universit\'e de Toulouse, 9 Avenue du Colonel Roche,
BP 44346, CEDEX 4, 31028 Toulouse, France}

\abstract{Increasing evidence suggests that cosmological sheets, filaments, and voids may be substantially magnetized today. The origin of magnetic fields in  the intergalactic medium (IGM) is, however, currently uncertain. It seems  well known that non-standard extensions to the physics of the standard model can provide mechanisms susceptible of magnetizing the universe at large. Perhaps less well known is the fact that standard, classical physics of matter--radiation interactions actually possesses  the same potential. We discuss a magnetogenesis mechanism based on the exchange of momentum between hard photons and electrons in an inhomogeneous IGM. Operating in the neighborhood of ionizing sources during the epoch of reionization, this mechanism is capable of generating magnetic seeds of relevant strengths over scales comparable to the distance between ionizing sources. In addition, summing up the contributions of all ionizing sources and taking into account the distribution of gas inhomogeneities, we show that this mechanism leaves the IGM, at the end of reionization, with a level of magnetization that might account, when amplification mechanisms take over, for the  magnetic fields strengths in the current cosmic web.}


\begin{document}


\section{Introduction}

From planets to galaxies through stars and the interstellar medium, the existence of magnetic fields in astrophysical objects and environments has been detected by various, complementary techniques (see, e.g.,  \cite{Widrow2002, Beck2013, Vallee2011a, Vallee2011}). On the scales of galaxy clusters too, there is growing evidence that the warm-hot, ionized gas of the intracluster medium is essentially always magnetized \cite{GOVONI2004a,Feretti2012,Han2017}, typically up to the level of a few microgauss. Magnetic fields may well be present even on the yet larger scales of intergalactic filaments ($\sim$10 nG typically) and cosmic voids (possibly $\ge$10$^{-18}$ G) as hinted at by observations of distant blazars \cite{Aleksic2010,Neronov2010,Tavecchio2011,Finke2015,Ackermann2018}.
In a word, magnetic fields seem ubiquitous in the universe.

While their existence within collapsed cosmic structures may be explained, at least to some degree, by the action of a local mechanism that generates fields with, to the least, a non-vanishing dispersion on some scales, followed by a sequence of various dynamo actions (see, e.g.,  \cite{Brandenburg2005, Schober2013, Schober2015, Federrath2016}), their origin on large, cosmological scales remains unclear. According to the current paradigm, they were  first generated as weak seeds
that were later amplified and reshaped, perhaps first on small scales  in the post-recombination intergalactic medium (IGM) through collisionless plasma instabilities (e.g.,  \cite{FalcetaGoncalves15}), and/or through adiabatic compression and various dynamo mechanisms during or after structure formation \cite{2009A&A...494...21A,Ryu2012,Donnert2018}. Had they been generated with strengths larger than a few nano-Gauss, magnetic fields would have noticeably affected subsequent structure formation \cite{Wasserman78,Kim96,Tashiro06,Marinacci2016,Varalakshmi2017}.

However, the origin of the seed magnetic fields themselves, particularly on the largest scales, is~unsettled, despite the many magnetogenesis mechanisms that have
been proposed in the literature (see for instance \cite{Widrow2002, Kulsrud08, Ryu2012, Widrow2012, Durrer2013} for reviews). Many of these (reviewed in \cite{Widrow2012,Durrer2013,Subramanian16}) rely on high energy, beyond the standard model physics possibly operating in the early universe. In the post-recombination
universe, classical plasma physics can also generate
magnetic field seeds through plasma instabilities (e.g.,  \cite{Gruzinov2001,Schlickeiser2003, Medvedev2006, Lazar2009, Bret2009,Schlickeiser2012}),
the Biermann \cite{Biermann1950} battery (e.g.,  \cite{Pudritz89, Subramanian1994, Ryu1998, Gnedin00,Naoz2013}), or the momentum transfer of photons or protons to electrons (e.g.,  \cite{Harrison1970,Mishustin1972,Harrison1973,Birk2002,Langer2003,Langer05,Fenu2011,Saga2015,Hutschenreuter2018}). Whether all these mechanisms are suitable candidates for explaining the origin of the magnetic fields probably permeating the cosmic web is debated, and will hopefully be answered in the near future notably thanks to the current and forthcoming large radio telescopes (see~\cite{Beck2015} and pages 369--597 of \cite{Bourke2015}).

We summarize here the basic features and results of an astrophysical mechanism, based on the photoionization of the IGM, that is bound to have contributed to the magnetization of the cosmic web during the first billion years of the universe, in the epoch of reionization. Its principles have been explored in \cite{Langer05,Durrive2015}, and the resulting, average strength of the magnetic field that permeates the universe by the end of reionization has been estimated in \cite{Durrive2017}.

\section{Outline of the Physical Mechanism}\label{sec:proc}
In principle, the ``recipe'' for the generation of magnetic fields is rather simple.~First, some~mechanism must spatially separate positive and negative electric charge carriers.~Second, this~charge separation must be sustained, at least effectively, so that the medium becomes permeated with a large-scale electric field. Third, for it to be suitable for the induction of magnetic fields, by~virtue of Faraday's law, this electric field must possess a curl. In the context of astrophysical plasmas, this~rotational electric field enters the magnetic field induction equation as a source term,
\begin{equation}\label{eq:induction}
 \partial_t \vec{B} = \vec{\nabla}\times\left(\vec{v}\times\vec{B}\right) - c \vec{\nabla} \times \vec{E}_\mathrm{s}
\end{equation}
(magnetic field diffusion has been omitted here). The question of astrophysical magnetogenesis thus essentially boils down to identifying the propitious epochs and environments for such rotational electric fields to appear. The Biermann battery \cite{Biermann1950} is perhaps the best known mechanism that creates rotational electric fields such that
\begin{equation}
 \vec{\nabla} \times \vec{E}_\mathrm{s} = \frac{\vec{\nabla}n_e \times\vec{\nabla}p_e }{en_e^2}
\end{equation}
where $p_e$ and $n_e$ are the electron pressure and number density, and $e$ is the elementary charge. In~the cosmological context, it has been explored by several authors \cite{Subramanian1994,Kulsrud1997,Gnedin00, Hanayama2005} who demonstrated that magnetic fields of strengths of $10^{-20}$--$10^{-18}$ Gauss, depending on the characteristic gradient scales involved, are thus generated in regions where the cosmological plasma is baroclinic, typically within ionization fronts during cosmological reionization or in shocks and caustics associated with structure formation by gravitational collapse.

Magnetizing the entire IGM is a little bit trickier. Of course, setting aside inflationary mechanisms or primordial universe phase-transitions, one may invoke the possibility that magnetic fields were first created within galaxies and subsequently injected into the circum- and inter-galactic media by galactic winds and AGN jets \cite{Rees1987,Daly1990,Kronberg1994b,Kronberg1999,Furlanetto2001,Hanasz2013,Beck13}. The resulting volume filling factor of these processes remains, however, highly uncertain. In addition, the origin of the initial magnetic seed fields generated within galaxies remains to be specified in that scenario, as well as  the efficiency of any amplification mechanism that is required to produce the strong enough fields that would be injected into the IGM.

\begin{figure}[H]
 \centering
 \includegraphics[width=0.70\textwidth]{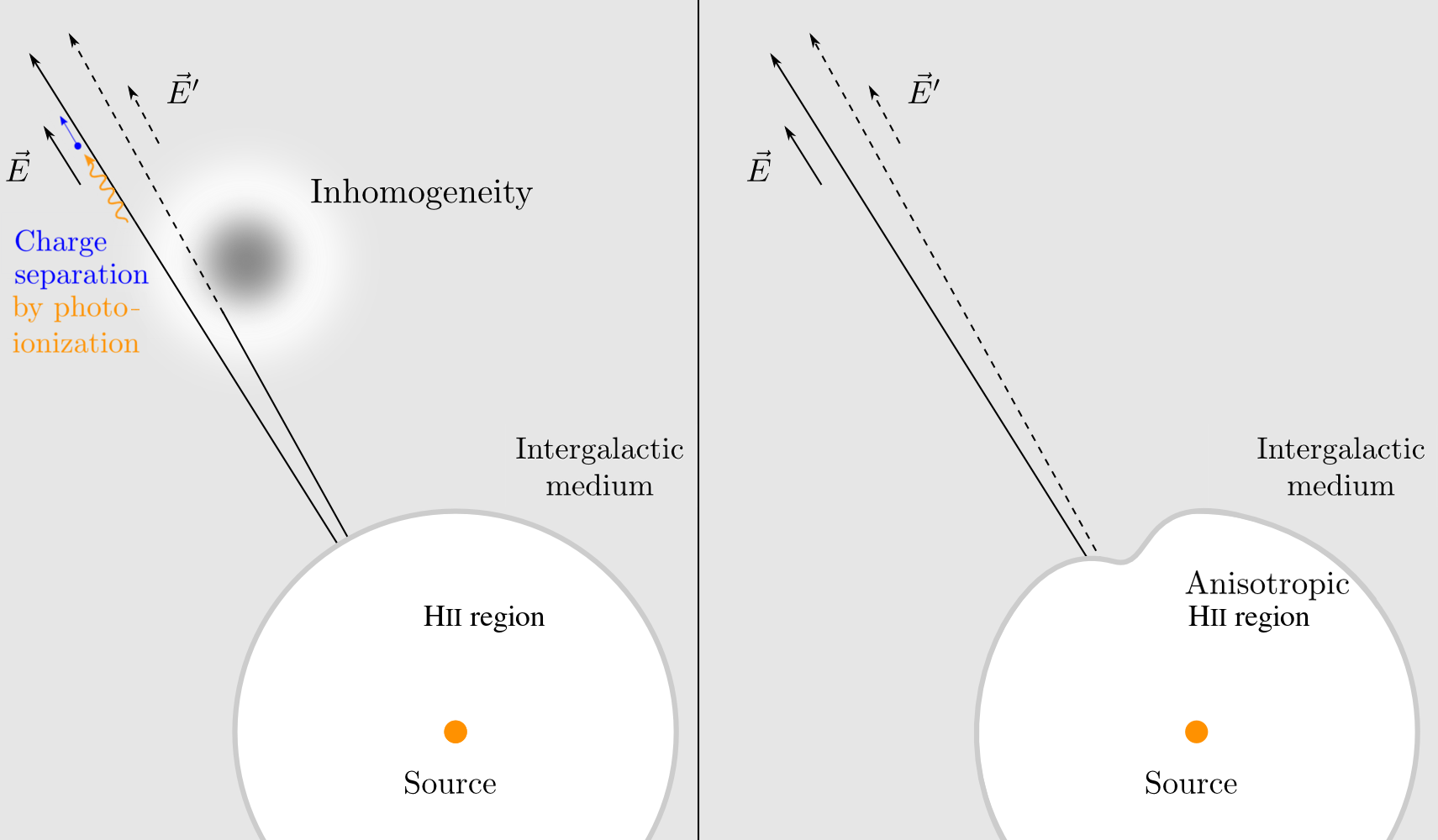}\\
 (\textbf{a})\hspace{150pt}(\textbf{b})
 \caption{Inhomogeneities at the origin of rotational electric fields. An ionizing source creates an H\textsc{ii} bubble around itself. Photons above the hydrogen ionization threshold may propagate deeper into the IGM, where, occasionally, they hit neutral atoms and eject new electrons. This process occurs continuously, and the resulting charge separation induces an electric field. This electric field may possess a curl either (\textbf{a}) because of gas inhomogeneities present in the IGM or (\textbf{b}) because of anisotropies in the bubble boundary that both induce differences in adjacent photon paths.\label{fig:aniso}}
\end{figure}

Cosmological reionization has most likely contributed to the global magnetization of the entire IGM. The first step of charge separation occurs naturally in that epoch. In fact, before the end of reionization, two different regions can be identified. First, the immediate surroundings of ionizing sources (first stars and galaxies, possibly quasars) are essentially fully ionized by photons with energies just at the threshold of hydrogen (and helium \textsc{i}) ionization forming cosmological H\textsc{ii} regions equivalent to Str\"omgren spheres. Higher energy photons can, however, penetrate beyond the Str\"omgren edge of such bubbles into the otherwise essentially neutral IGM. There, occasionally, they hit neutral atoms and eject new electrons. As long as the ionizing source shines, the resulting charge separation gives birth to an electric field. Now, if the distribution of matter around ionizing sources were completely isotropic, the electric field would obviously be curl-free. Nevertheless, two circumstances ensure that local isotropy is broken (see Figure \ref{fig:aniso}). First, the IGM is naturally inhomogeneous, if only because of the primordial density fluctuations initially present at all scales. As we discuss     below, the most favorable inhomogeneities are only mildly nonlinear, but any overdensity (underdensity) locally enhances (lessens) the photoionization process, meaning that behind it the strength of the resulting electric field is smaller (larger) than it is 
 along photons trajectories that miss density contrasts. Thus, the electric field varies \emph{across} the photon radial trajectories and possesses a curl. Second, due to the presence of inhomogeneities, the H\textsc{ii} bubbles themselves are strongly aspherical (see for \mbox{instance~\cite{Ciardi2001, Pawlik2017}}), which contributes to making  the flux of hard photons that escape into the IGM~anisotropic.

In \cite{Durrive2015}, we analyzed in detail the mechanism by which non-zero magnetic fields emerge in the neutral surroundings of ionizing sources during cosmological reionization.~At the fundamental level, it~relies on the momentum transfer between hard photons and initially bound electrons. It may in fact be  described microscopically as a perturbation of the electron distribution function in the IGM. We~then reduced the description to that of a single fluid at a macroscopic level and obtained a generalized Ohm's law whose curl leads to the induction Equation (\ref{eq:induction}), where, this time, the source electric field contains an additional term such that
\begin{equation}\label{eq:newinduction}
 \vec{\nabla} \times \vec{E}_\mathrm{s} =   \frac{\vec{\nabla}n_e \times\vec{\nabla}p_e }{en_e^2} + \vec{\nabla}\times\left(\frac{\dot{\vec{p}}}{e n_e}\right)
\end{equation}
where $\dot{\vec{p}}\equiv \dot{\vec{p}}(\vec{r})$ is the rate of electron momentum creation at position $\vec{r}$ in the IGM due to the higher energy photons. As we have shown in \cite{Durrive2015}, this new source term depends notably on the neutral gas density, the spectrum of the ionizing source, the attenuation along the photon trajectories, and~the distance from the edge of the H\textsc{ii} bubble (a geometric dilution of photons). By  the integration of Equation~(\ref{eq:induction}), keeping only the photoionization source term of Equation (\ref{eq:newinduction}), we were able to obtain the general expression for the magnetic field thus induced (not considering the role of the asphericity of ionized bubbles), and we examined, in a simple model of the cosmological reionization context, the~typical spatial distribution and strengths of the magnetic fields obtained.

\section{Resulting Magnetic Fields}

\begin{figure}[H]
 \centering
 \begin{minipage}{.3\textwidth}
  \centering
  \includegraphics[width=\textwidth]{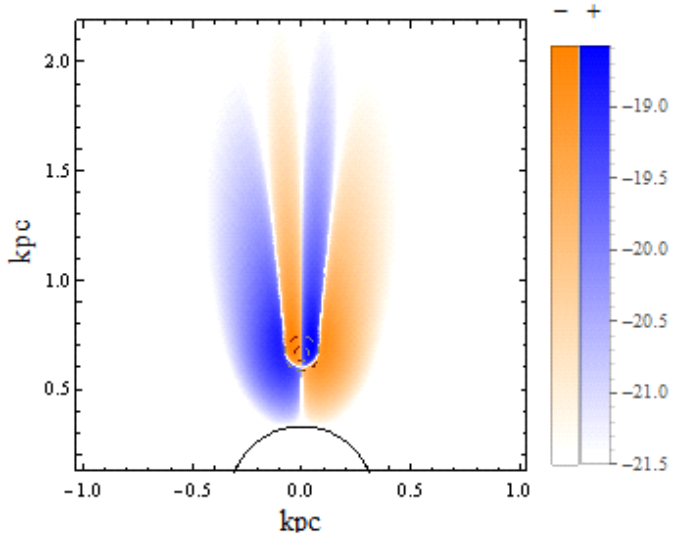}
 \end{minipage}
 \begin{minipage}{.3\textwidth}
  \centering
  \includegraphics[width=\textwidth]{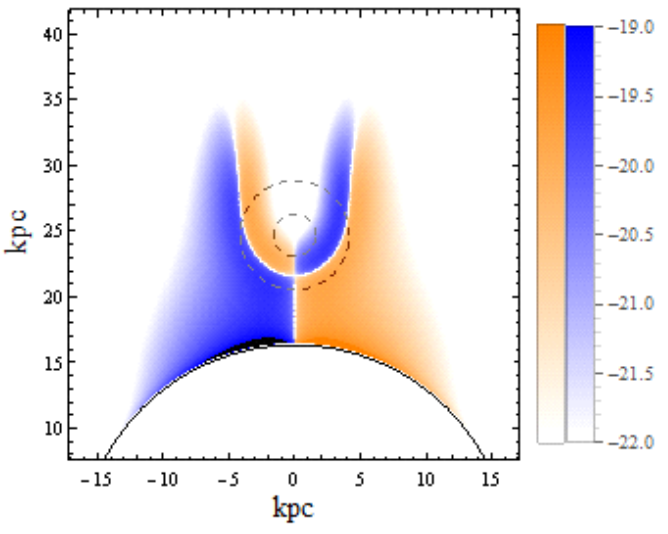}
 \end{minipage}
 \begin{minipage}{.3\textwidth}
  \centering
  \includegraphics[width=\textwidth]{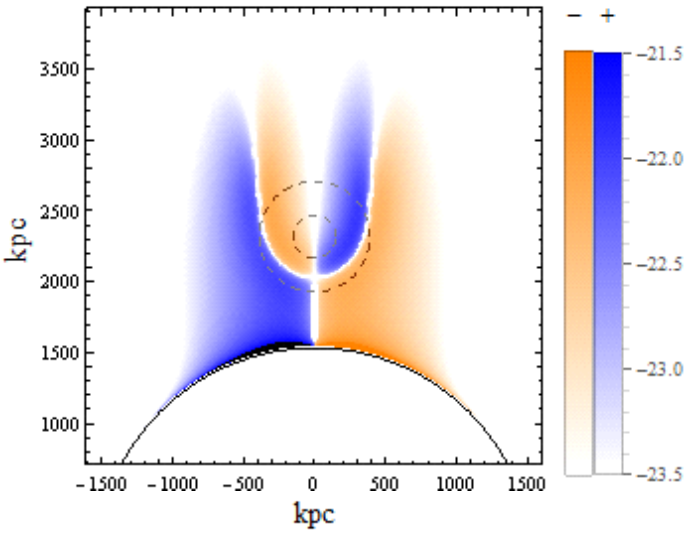}
 \end{minipage}
 \caption{Strength and spatial configuration of the magnetic fields induced by the radiation of (\textbf{a}) Population III stars at $z$$\sim$20, (\textbf{b}) a primeval galaxy at $z$$\sim$10, and (\textbf{c}) a quasar at $z$$\sim$10. In each panel, the blue areas correspond to the magnetic field pointing toward the reader, orange in the opposite direction. This peculiar geometrical configuration of the fields simply reflects the axisymmetry of the matter distribution with respect to the ionizing source. Strengths are in Gauss in logarithmic scale. The~solid line arc represents the edge of the H\textsc{ii} region created by the ionizing source (not shown). Inner and outer dashed black circles delineate, respectively, the one sigma location of the overdensity and of the compensating underdensity (the latter being responsible for the reversal of the magnetic field).  Note that the scales change from panel to panel. \label{fig:local}}
\end{figure}

Depending on their spectrum, luminosity, and lifetime, different sources generate different magnetic fields at different epochs. As illustrative numerical applications, we considered three different types of sources that presumably have contributed to reionization: clusters of  Population III stars, primeval galaxies, and high redshift quasars. We modeled an inhomogeneity in the IGM by a Mexican hat density profile in which a central overdensity is compensated by a surrounding underdensity (see Figure \ref{fig:aniso}a). The most favorable gas inhomogeneities in the IGM are mildly nonlinear, $\delta\rho/\rho$$\sim$5, corresponding to the density contrast of cosmic structures that are in the turn-around phase of their gravitational collapse. Much smaller inhomogeneities are not contrasted enough to make adjacent photon trajectories significantly different. Strongly nonlinear density contrasts are too opaque  and possess much sharper, narrower boundaries that limit the extent of the areas over which rotational electric fields; hence, 
  magnetic fields are induced. Finally, the strength of the magnetic field, growing linearly with time (see \cite{Langer05,Durrive2015}), is shown in Figure \ref{fig:local} for each ionizing source type, assuming a source lifetime of $100$ Myr.

Population III star clusters generate relatively stronger fields (up to $\sim$$10^{-18}$ G)  on short distances (1--2 kpc). Assuming that they formed by molecular cooling in gas overdensities of mass $10^6\, M_\odot$ \cite{Barkana2001a}, their half physical mean separation is roughly 10 kpc, estimated from the abundance of their $\sim$$3\sigma$ parent halos \cite{Mo2002a}. In addition, the size of their H\textsc{ii} bubbles is of the order of a fraction of kpc. It thus seems that these sources magnetize essentially their immediate neighborhood, leaving an important fraction of the IGM unmagnetized. Luminous quasars, on the contrary, magnetize less (up to $\sim$$10^{-21}$~G) but over much larger distances (several Mpc). Considering that these sources, rare at $z=10$, are~hosted by $5 \sigma$ halos, their half physical mean separation is of a couple of Mpc.  Their H\textsc{ii} bubbles have radii of the order of the Mpc, which means that these sources emit photons of large enough mean-free paths to contribute to globally magnetizing the IGM. Finally, primeval galaxies combine modestly high amplitudes (up to $\sim$$10^{-19}$ G) and reasonably large scales (tens of kpc). Assuming they are hosted in $10^8\, M_\odot$ halos, their half physical mean separation is typically several tens of kpc, while the radius of their H\textsc{ii} bubbles is of the order of a few tens of kpc. We thus may conclude that such sources have probably magnetized the entire IGM before the end of reionization.

\section{Average Magnetic Energy Density Seeded in the IGM}
To support this conclusion, we estimated in \cite{Durrive2017} the level of global magnetization produced in the universe by the end of reionization by the mechanism outlined above. The final result naturally depends on the distribution of all  ionizing sources, their spectral properties, the epochs at which they start shining, and the shape and distribution of density inhomogeneities in the IGM. In \cite{Durrive2015}, we~derived a very detailed formula giving the strength of the magnetic field generated within and around a given overdense cloud in the neighborhood of an ionizing source. The latter expression is rather involved but, fortunately, we were also able to identify the characteristic length scales of the problem useful for modeling simply the magnetized area around  the  given overdensity.

Building upon this, we employed  in \cite{Durrive2017} the Press–Schechter (PS) formalism \cite{Press1974b} to estimate the statistical distribution of ionizing sources and of overdensities around the sources.~To be more precise, we~considered dark matter (DM) halos that are massive enough to host luminous sources and DM overdensity regions that have not yet collapsed but that contain diffuse baryonic overdense clouds. Then, using an approximate expression for the magnetic field generated around overdensities, we~estimated the magnetic field generated by all   sources forming during the epoch of reionization. In~fact, for simplicity, we concentrated only on the contribution of faint, primeval galaxies.~Indeed, the~latter are probably the dominant contributors to the reionization process itself, as has been suggested in the literature (see for instance \cite{McQuinn2016b} and references therein).

\begin{figure}[H]
 \centering
 \begin{minipage}{.47\textwidth}
  \centering
  \includegraphics[width=.97\linewidth]{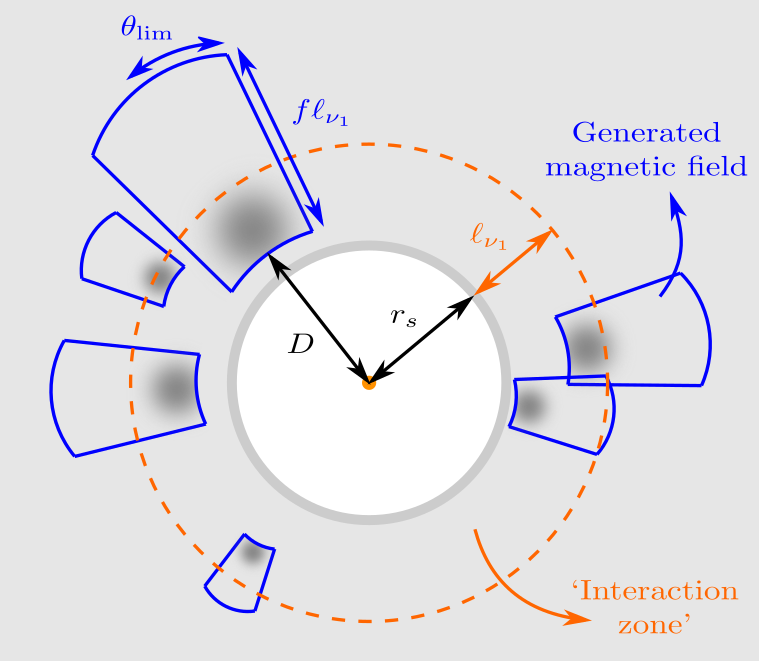}
 \end{minipage}
 \begin{minipage}{.47\textwidth}
  \centering
  \includegraphics[width=1.03\linewidth]{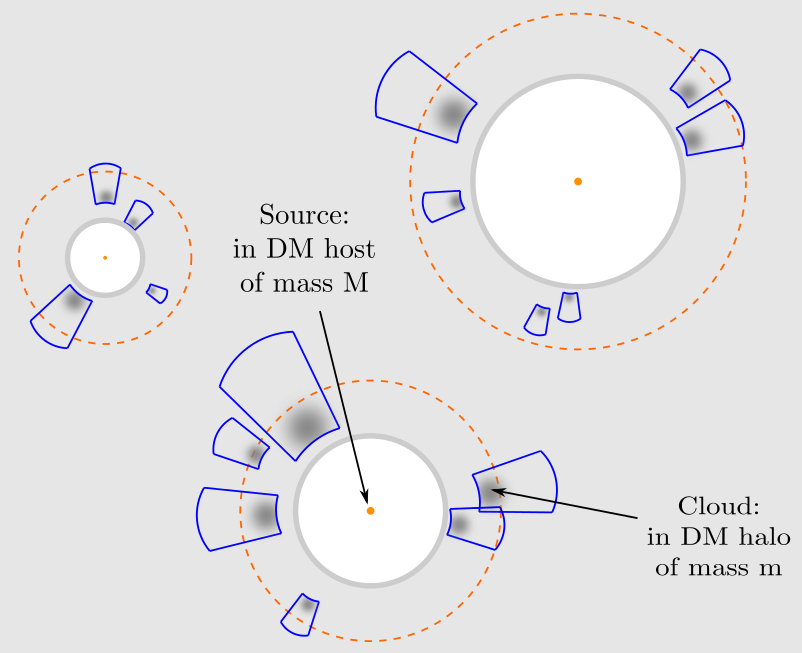}
 \end{minipage}%
 \caption{Illustration of the approach we followed for estimating the global magnetization level of the IGM. We first (\textbf{a}) evaluated the magnetic field generated around one source, surrounded by a distribution of clouds, and then (\textbf{b}) considered a distribution of sources. The orange dot represents an ionizing source. Gray spots represent small baryonic inhomogeneities in the IGM. Blue frames delineate the characteristic regions that are magnetized around 
  such gas clouds.
 }
 \label{fig:ExampleOfResultsVaryRhob}
\end{figure}

The details of our calculations are exposed in \cite{Durrive2017}. Here, we simply outline our approach, illustrated in Figure \ref{fig:ExampleOfResultsVaryRhob}, which consists in the following steps.

\begin{enumerate}[leftmargin=*,labelsep=4.9mm]
 \item First, along the lines outlined in Section \ref{sec:proc}, we considered an isolated source and a gas inhomogeneity in its vicinity. We then condensed all the details into a simpler and more tractable version of the results we obtained in \cite{Durrive2015}. We then derived an efficient expression for the magnetic energy density $E_m(D)$ associated with any cloud of mass $m$ at a given distance $D$ of the ionizing~source.
 \item Second, we summed up the contributions of all the clouds surrounding the source. For this, we needed to estimate the distribution of such baryonic clouds, which we did by considering their underlying DM halos. Using the PS formalism, we obtained that the source, contained in a DM halo of mass $M$, contributes to the magnetization of the IGM by injecting a magnetic energy~density
       \begin{equation}
        E_M = \int_{r_s}^{r_s + \ell_{\nu_1}} \int_{m_\text{min}}^{m_\text{max}}\, E_m(D)\,\dd^2 P(D, m|M)
        \label{eq:E_M}
       \end{equation}
       where $\dd^2 P(D, m|M)$ is the probability of finding a DM cloud of mass $m$ within a spherical shell of volume $4 \pi D^2 \dd D$, at a distance $D$ from the halo of mass $M$. As one can see from the boundaries of the first integral in Equation (\ref{eq:E_M}), we take into account only the clouds outside the H\textsc{ii} bubbles (of radius $r_s$), but not too far from it, namely within an ``interaction zone'' set by the photoionization mean free path $\ell_{\nu_1}$ beyond which  the mechanism is no longer significantly efficient (orange dashed lines in Figure \ref{fig:ExampleOfResultsVaryRhob}). The boundaries of the second integral, $m_\mathrm{min}$
       and $m_\mathrm{max}$ for the mass of the clouds, are important parameters of this process, and we discuss in details how we chose them in \cite{Durrive2017}.
 \item Third, we considered the full cosmological context, as illustrated in the right panel of Figure~\ref{fig:ExampleOfResultsVaryRhob}. The energy density $E_M$ generated around each source must be integrated over the distribution of DM halos containing the sources. However, ionized bubbles of individual sources start to overlap toward the end of the epoch of reionization. Since the mechanism outlined in Section \ref{sec:proc} takes place only in the neutral IGM, the efficiency of magnetic field generation actually decreases as reionization proceeds. Hence, we weighed the contribution of all sources by a factor $1-Q_i(z)$ (where $Q_i(z)$ is the volume filling factor of ionized bubbles at redshift $z$) when summing them up in order to reduce the generated magnetic field energy as time increases according to the ionization of the IGM. We thus estimated the mean physical magnetic energy density generated by photoionizations during the epoch or reionization as
       \begin{equation}
        \frac{B_\mathrm{p}^2(z)}{8 \pi} = (1+z)^4 \int^{z_0}_z \dd z' \frac{1-Q_i}{H\,(1+z')^5}\int^{M_\mathrm{max}}_{M_*} \dd M\, E_M\, g_\mathrm{gl}~\frac{\dd n_M}{\dd M}
        \label{eq:b2pz}
       \end{equation}
       where $\frac{\dd n_M}{\dd M}$ is the mass function of the DM halos, with mass $M$, hosting the sources. We also used the PS formalism to compute the latter, and we discuss our choices for the values of the boundaries $M_*$ and $M_\mathrm{max}$ in \cite{Durrive2017}. The parameter $z_0$ is the redshift at which the first sources form. We also introduced the rate $g_\mathrm{gl}$ at which sources in DM halos switch on, with a numerical value chosen in order for our model to be consistent with recent observational constraints on reionization (see \cite{PlanckCollaboration2016}). Finally, the presence of the Hubble parameter $H$ and of the $(1+z)$
       factors reflects the expansion of the universe that dilutes the magnetic fields.

\end{enumerate}

\begin{figure}[H]
 \begin{center}
  \begin{minipage}{.49\textwidth}
   \includegraphics[width=0.9\textwidth]{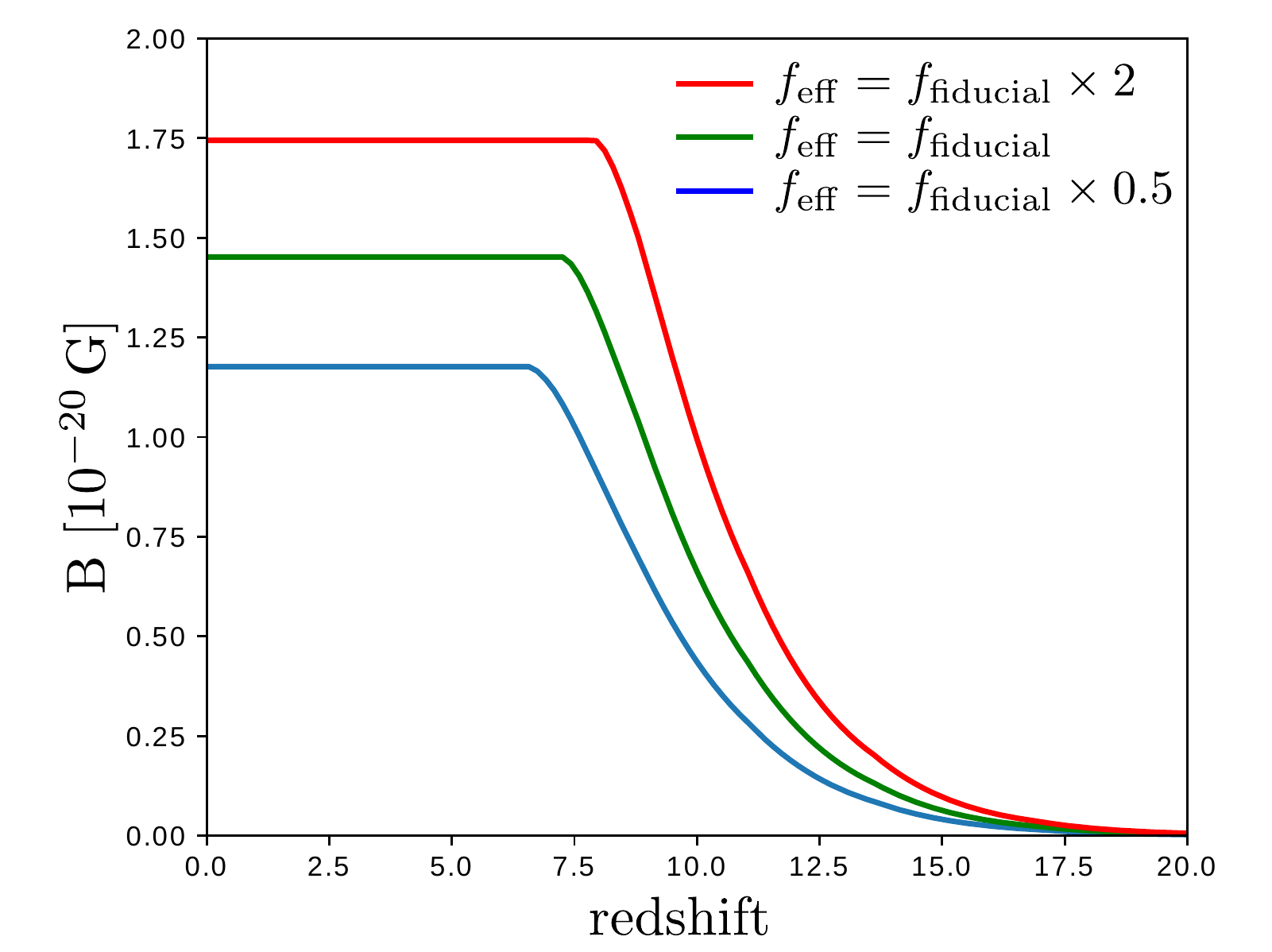}
  \end{minipage}
  \begin{minipage}{.49\textwidth}
   \includegraphics[width=0.9\linewidth]{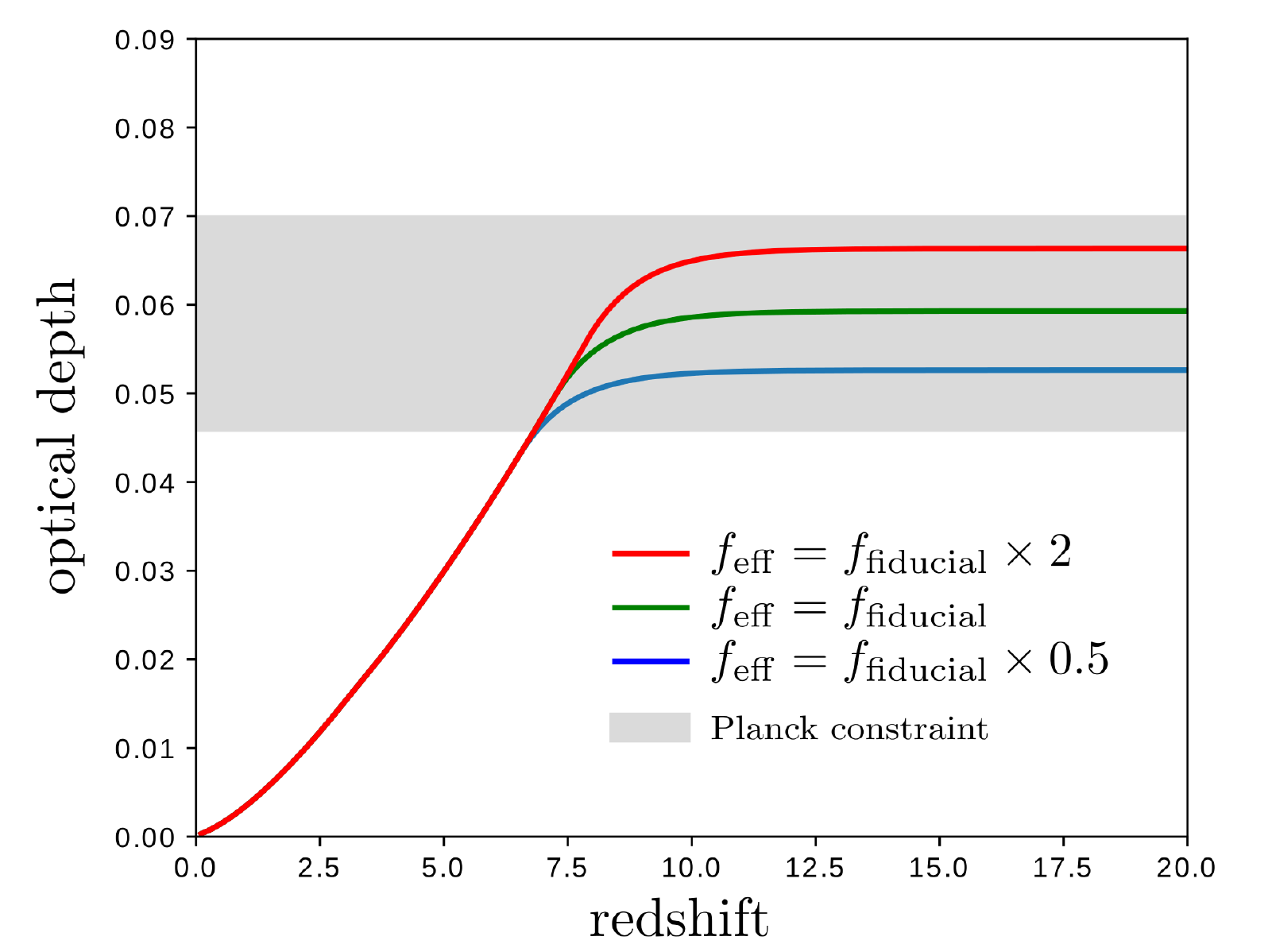}
  \end{minipage}%
  \caption{(\textbf{a}) Evolution with redshift of the mean comoving magnetic field strength accumulated in the IGM in different reionization histories. The green curve corresponds to the fiducial model assumed in \cite{Durrive2017} where $f_\mathrm{eff}\equiv f_*f_\mathrm{esc}$ is an effective reionization efficiency. It combines the fraction $f_*$ of baryons converted into stars in the halos hosting the ionizing sources and the fraction $f_\mathrm{esc}$ of ionizing radiation that actually escapes into the IGM. The blue and red curves correspond, respectively, to reionization scenarios that are half and twice as efficient. All three considered reionization histories  are in agreement with the Planck constraints ($\tau = 0.058 \pm 0.012$, \cite{PlanckCollaboration2016}), as illustrated in panel
  (\textbf{b}), which shows the cumulative Thomson optical depth $\tau$ to the CMB.}
  \label{fig:bcom}
 \end{center}
\end{figure}

In Figure \ref{fig:bcom}, for illustration, we plot the comoving strength of the generated magnetic field $B_\mathrm{c} = B_\mathrm{p}/(1 + z)^2$, as a function of redshift, in three different reionization histories, all consistent with the value of the Thomson optical depth to the cosmic microwave background measured by the Planck collaboration \cite{PlanckCollaboration2016}.
The global shape of the curves is naturally understood as follows. Above $z = 20$, there are no galaxies, and the magnetic field strength is equal to zero. As time passes, galaxies form and their radiation induces magnetic fields, i.e.\ a small fraction of their radiation energy is converted into magnetic energy that accumulates in the IGM. Once the universe is totally ionized, the mechanism outlined above stops, and a plateau is reached (in comoving units). Note that in physical units, the strength of the magnetic field permeating the universe by the end of reionization due to this process is a few $10^{-18}$ Gauss, which is a suitable seed value for any succeeding amplification by nonlinear~processes.

\section{Discussion}

The model we summarized here is arguably simplistic in some aspects and can be improved in several ways. In particular, in the estimate of the average magnetization of the universe, we considered only gas overdensities, and neglected the contribution of underdense regions (see Section \ref{sec:proc}), which could, in principle, multiply the result obtained above by a factor of two. Similarly, we did not take into account the asphericity of the cosmological H\textsc{ii} regions, which would probably be best investigated through numerical simulations.

Finally, we neglected a few time-dependent aspects in this first approach. Indeed, we assumed for instance that the H\textsc{ii} regions surrounding the ionizing sources have reached their steady state. However, in the transitory phase during which they grow, hard photons reaching the neutral IGM are less diluted, so that the mechanism is more efficient and
  initially closer to the photon sources. At~the same time, it is slightly less efficient and 
  further away from the sources, since photons above the hydrogen ionization threshold are more absorbed by the intervening matter. Whether taking this into account would increase or decrease the overall mean generated magnetic field is not obvious. Note,~however, that nonlinearities develop in the cosmic velocity field as structure formation proceeds. They~enter into play in Equation (\ref{eq:induction}) basically when the seed magnetic field has reached its final strength at a given scale~\cite{Langer05}. We also know from numerical simulations that
strong turbulent motions arise naturally during the formation of the first galaxies \cite{Greif2008,Sur2012}. In addition, mechanical feedback associated to the formation and evolution of large-scale structure after reionization injects turbulence into the IGM (e.g.,  \cite{Rauch2001,Ryu2008,Ravi2016,Donnert2018}). Magnetic field amplification by shear and adiabatic contraction thus sets in early on, counterbalancing (and overwhelming) the effect of the dilution due to the expansion of the universe, at least within the nodes, filaments, and sheets of the cosmic web. The strength shown in Figure~\ref{fig:bcom} is thus likely an underestimate of the actual magnetization level of these structures by the end of reionization. Within cosmic voids, collisionless plasma instabilities have the potential to amplify rapidly the magnetic seed fields \citep{FalcetaGoncalves15} and bring them to, and maintain them above,
  the lower limits suggested by the observation of distant blazars and cited in the introduction.

As a final note, let us point out that the many magnetogenesis mechanisms proposed in the literature are  not mutually exculsive. Quite on the contrary, several may contribute either simultaneously or sequentially, sympatrically or allopatrically, to the magnetization of the IGM\footnote{Among the possible contributions is the resistive mechanism suggested in \cite{Miniati2011}, in which \textit{temperature} inhomogeneities of the IGM are responsible for spatial variations of the resistivity that induce rotational electric fields thanks to cosmic ray driven return currents.}. Disentangling their respective contributions may become possible in the near future thanks to synchrotron measurements and Faraday tomography (e.g.,  \cite{Akahori2014b,Akahori2018})---especially when compared to numerical simulations that are able to track the various contributions \cite{Vazza2017,Katz2018}---or even in the statistical properties of the neutral gas during the epoch of reionization \cite{Venumadhav2017,Gluscevic2017}.

\vspace{6pt}



\authorcontributions{Conceptualization, Mathieu Langer and Jean-Baptiste Durrive; Formal Analysis, Jean-Baptiste Durrive and Mathieu Langer; Investigation, Jean-Baptiste Durrive and Mathieu Langer; Writing–Original Draft Preparation, Mathieu Langer and Jean-Baptiste Durrive; Writing–Review \& Editing, Mathieu Langer.}

\funding{This work has been in part supported by MEXT Grant-in-Aid for Scientific Research on Innovative Areas No. 15K2173 (J.B.D.). M.L.
 acknowledges partial financial support from the ``Programme National de Cosmologie \& Galaxies'' (PNCG) funded by CNRS/INSU-IN2P3-INP, CEA and CNES, France.}

\acknowledgments{The authors very gratefully acknowledge their fruitful collaboration with H. Tashiro (supported in part by JSPS Grants-in-Aid for Scientific Research under Grant No. 5K17646) and N. Sugiyama (supported in part by JSPS Grants-in-Aid for Scientific Research under Grant Nos. 25287057 and 15H05890) that led to  the publication of \cite{Durrive2017}. They also would like to warmly thank all the organizers of the conference ``The Power of Faraday Tomography: Towards 3D Mapping of Cosmic Magnetic Fields'' for giving them the chance to present these results. Finally, they thank the referees for their careful reading, their comments, and their suggestions.}

\conflictsofinterest{The authors declare no conflict of interest.}

\reftitle{References}



\end{document}